\input{aipcheck}

\documentclass[
    ,final            
  ]
  {aipproc}

\layoutstyle{6x9}

\usepackage{amssymb,amsmath}

\begin{document}

\title{Direct-Photon Production in Au+Au Collisions \\at RHIC}


\author{Klaus Reygers for the PHENIX Collaboration}{
  address={University of M{\"u}nster, Institut f{\"u}r Kernphysik, \\
           Wilhelm-Klemm-Stra{\ss}e 9, 48149 M{\"u}nster, Germany}
}



\begin{abstract}
  Results from the PHENIX experiment on direct-photon production in
  Au+Au collisions at $\sqrt{s_\mathrm{NN}} = 200$\;GeV for transverse
  momenta $1 \lesssim p_\mathrm{T} \lesssim 13$\;GeV/$c$ are
  presented. Direct-photon yields at high $p_\mathrm{T}$ scale as
  expected for particle production in hard processes. This supports
  jet-quenching models which attribute the suppression of
  high-$p_\mathrm{T}$ hadrons to the energy loss of fast partons in
  the quark-gluon plasma.  The low-$p_\mathrm{T}$ direct-photon
  spectra, measured via $e^+e^-$ pairs with small invariant masses,
  are possibly related to the production of thermal direct photons in
  Au+Au collisions at RHIC.
\end{abstract}

\maketitle


\section{Introduction}
Depending on their transverse momenta, $p_\mathrm{T}$, direct photons,
{\it i.e.}, photons not coming from hadron decays like $\pi^0
\rightarrow \gamma \gamma$, convey information about different aspects
of ultra-relativistic nucleus-nucleus (A+A) collisions.

Direct photons at high $p_\mathrm{T}$ are produced in initial
parton-parton scatterings with large momentum transfer (hard
scatterings) in processes like quark-gluon Compton scattering
($\mathrm{q}+\mathrm{g} \rightarrow \mathrm{q}+\gamma$) and
quark-antiquark annihilation ($\mathrm{q}+\bar{\mathrm{q}} \rightarrow
\mathrm{g}+\gamma$). Unlike scattered quarks and gluons, photons from
initial hard scatterings are not affected by the hot and dense medium
subsequently produced in a A+A collision. High-$p_\mathrm{T}$ direct
photons ($p_\mathrm{T} \gtrsim 6$\;GeV/$c$ in Au+Au collisions at
$\sqrt{s_\mathrm{NN}} = 200$\;GeV) can therefore be used to test the
theoretical description of the initial phase of a A+A collision with
perturbative Quantum Chromodynamics (pQCD). Moreover, they serve as a
measure of the rate of initial hard parton-parton scatterings in A+A
collisions.

A significant fraction of low-$p_\mathrm{T}$ direct photons ($1
\lesssim p_\mathrm{T} \lesssim 3$\;GeV/$c$ in Au+Au collisions at
$\sqrt{s_\mathrm{NN}} = 200$\;GeV \cite{Turbide:2003si}) is expected
to come from the thermalized medium of deconfined quarks and gluons,
the quark-gluon plasma (QGP), possibly created in A+A collisions.
These so-called thermal photons carry information about the initial
temperature of the QGP. The QGP created in a A+A collision expands and
cools until the critical temperature for the transition to a hadron
gas is reached.  Thermal direct photons are created in the QGP as well
as in the hadron gas over the entire lifetime of these phases. The
initial temperature $T_\mathrm{i}$ of the QGP just after
thermalization can be extracted by comparing thermal photon data with
predictions from models which convolve thermal photon production rates
with the space-time evolution of the QGP and hadron gas phase.

At low and intermediate $p_\mathrm{T}$ (up to $p_\mathrm{T} \sim
6$\;GeV/$c$ in Au+Au collisions at $\sqrt{s_\mathrm{NN}} = 200$\;GeV)
the interaction of quarks and gluons from hard scattering processes
with the QGP might be a further significant source of direct photons
\cite{Turbide:2005fk}. An example for such a hard+thermal interaction
is a jet-photon conversion like
$\mathrm{q_\mathrm{hard}}+\mathrm{g_\mathrm{QGP}} \rightarrow
\mathrm{q}+\gamma$ in which the photon usually obtains a large
fraction of the momentum of $\mathrm{q_\mathrm{hard}}$.



\section{High-$p_\mathrm{T}$ Direct Photons}
In the PHENIX experiment at RHIC photons are measured with two types
of highly segmented electromagnetic calorimeters: a lead scintillator
sampling calorimeter (PbSc) and a lead glass Cherenkov calorimeter
(PbGl) \cite{Adler:2005ig}. Neutral pions and $\eta$ mesons are
reconstructed via their two-photon decay branch. The $p_\mathrm{T}$
spectrum of direct photons is obtained by subtracting the spectrum of
decay photons calculated based on the measured $\pi^0$ and $\eta$
spectra from the $p_\mathrm{T}$ spectrum of all photons.

The yield of direct photons from hard scattering processes in A+A
collisions relative to p+p collisions is expected to scale with the
increase of the initial parton luminosity per collisions.  This
increase is quantified with the nuclear overlap function
$T_\mathrm{AA}$.  In the absence of nuclear effects the nuclear
modification factor
\begin{equation}
R_\mathrm{AA}(p_\mathrm{T}) = 
  \frac{\mathrm{d}N/\mathrm{d}p_\mathrm{T}|_\mathrm{A+A}}
       {\langle T_\mathrm{AA}\rangle 
        \times \mathrm{d}\sigma/\mathrm{d}p_\mathrm{T}|_\mathrm{p+p}}
\end{equation} 
is unity for particles production in hard scattering processes.

\begin{figure}[tb]
\includegraphics[height=.35\textheight]{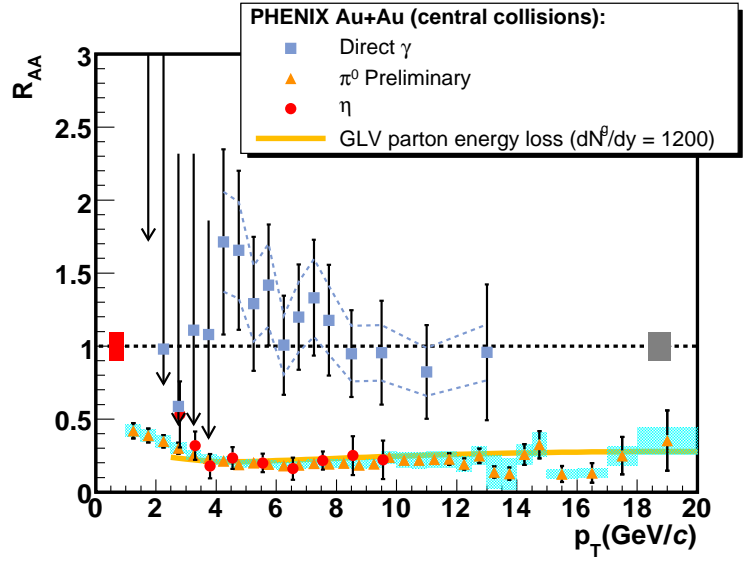}
\caption{Nuclear modification factor $R_\mathrm{AA}$ for direct photons,
  neutral pions, and $\eta$ mesons in central Au+Au collisions at
  $\sqrt{s_\mathrm{NN}} = 200$\;GeV. The suppression of $\pi^0$ and
  $\eta$ production can be described by energy loss of partons in
  the QGP.}
\label{fig:raa}
\end{figure}
Figure~\ref{fig:raa} shows the nuclear modification factor for direct
photons, neutral pions, and $\eta$ mesons in central Au+Au collisions
at RHIC. $\pi^0$ and $\eta$ production at high $p_\mathrm{T}$ is
suppressed by a factor of $\sim 5$ whereas high-$p_\mathrm{T}$
direct-photon yields scale as expected for particle production in hard
processes.  This is in line with jet-quenching models which assume
that the rate of hard processes in A+A collisions scales with
$T_\mathrm{AA}$ and which attribute the hadron suppression to the
energy loss of quarks and gluons from hard scattering in the
quark-gluon plasma.

\section{Low-$p_\mathrm{T}$ Direct Photons}
At low $p_\mathrm{T}$ ($\lesssim 4$\;GeV/$c$) the direct-photon signal
is small so that the standard method of measuring direct photons with
the electromagnetic calorimeters essentially only yields upper limits.
Therefore a new method based on the measurement of $e^+e^-$ pairs
identified with the aid of the Ring Imaging Cherenkov detector of the
PHENIX experiment was employed. The idea is that all sources of real
photons also produce virtual photons which decay into $e^+e^-$ pairs
with small invariant masses. An example is the $\pi^0$ Dalitz decay
$\pi^0 \rightarrow \gamma e^+e^-$.

\begin{figure}[th]
\includegraphics[height=.3\textheight]{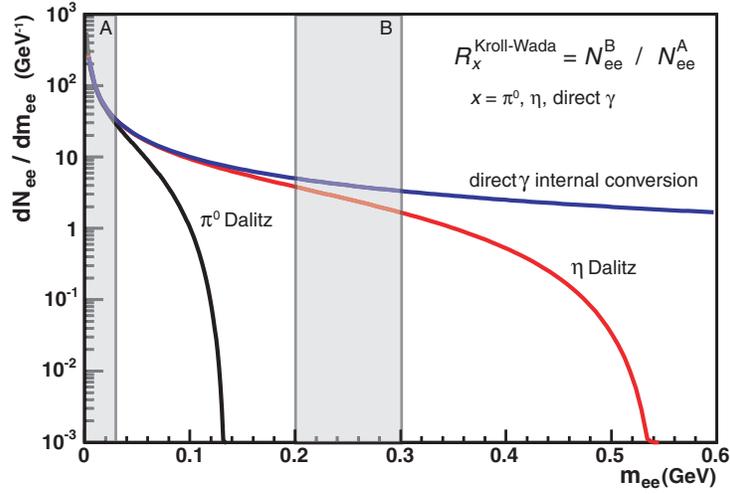}
\caption{Distribution of the invariant masses $m_{ee}$ of $e^+e^-$ pairs
  from $\pi^0$ and $\eta$ Dalitz decays and from the decay of virtual
  direct photons with $m_{ee} \ll p_\mathrm{T}^{ee}$ as given by
  Eq.~\ref{eq:kw}. The ratio $R_\mathrm{hadron}^\mathrm{Kroll-Wada}$
  of the $e^+e^-$ yields in the $m_{ee}$ intervals B and A expected
  from $\pi^0$ and $\eta$ decays is calculated as the weighted average
  of $R_{\pi^0}^\mathrm{Kroll-Wada}$ and
  $R_\eta^\mathrm{Kroll-Wada}$.}
\label{fig:kroll-wada}
\end{figure}

The basic assumption in this analysis is that the ratio of real direct
to all real photons is equal to the same ratio for virtual photons
with small invariant masses ($m_{\gamma^*} < 30$\;MeV):
\begin{equation}
\frac{\gamma_\mathrm{direct}}
     {\gamma_\mathrm{incl}}
=
\frac{\gamma_\mathrm{direct}^{*}}
     {\gamma_\mathrm{incl}^{*}}
\equiv
\frac{\gamma_\mathrm{direct}^{*\mbox{, } m < 30\;\mathrm{MeV}}}
     {\gamma_\mathrm{incl}^{*\mbox{, } m < 30\;\mathrm{MeV}}}\;.
\end{equation}
Furthermore, it is assumed that the distribution of the $e^+e^-$
invariant masses can be described by the Kroll-Wada formula
\cite{KrollWada}
\begin{equation}
  \frac{1}{N_\gamma} \frac{\mathrm{d}N_{ee}}{\mathrm{d}m_{ee}} = 
   \frac{2\alpha}{3\pi}
   \sqrt{1 - \frac{4m_e^2}{m_{ee}^2}} (1 + \frac{2m_e^2}{m_{ee}^2}) 
   \frac{1}{m_{ee}} |F(m_{ee}^2)|^2 (1 - \frac{m_{ee}^2}{M^2})^3 \; 
\label{eq:kw}
\end{equation}
which is shown in Fig.~\ref{fig:kroll-wada} for $e^+e^-$ pairs from
$\pi^0$ and $\eta$ Dalitz decays and from the decay of virtual direct
photons.  For $e^+e^-$ pairs from $\pi^0$ and $\eta$ Dalitz decays the
yield is suppressed towards higher $m_{ee}$ due to the mass $M$ of the
parent meson in the phase space factor $(1 - m_{ee}^2/M^2)^3$ whereas
no such suppression occurs for $e^+e^-$ pairs from virtual direct
photons as long as $m_{ee} \ll p_\mathrm{T}^{ee}$. For the small
$e^+e^-$ invariant masses considered here the form factor
$|F(m_{ee}^2)|$ is assumed to be unity in all cases.

The key idea is to make use of the increased signal/background ratio
for $e^+e^-$ pairs from virtual direct photons at higher $m_{ee}$. The
experimentally observed quantity is the ratio $R_\mathrm{data}$ of the
$e^+e^-$ pair yield at low invariant masses and at higher invariant
masses, {\it e.g.}, $R_\mathrm{data} = N_{ee}^{m < 30\;\mathrm{MeV}} /
N_{ee}^{200 < m < 300\;\mathrm{MeV}}$. In the absence of a direct
photon signal $R_\mathrm{data} =
R_\mathrm{hadron}^\mathrm{Kroll-Wada}$, {\it i.e.}, $R_\mathrm{data}$
can be calculated from Eq.~\ref{eq:kw} based on the known ratio
$\eta/\pi^0 = 0.45 \pm 0.1$ and the known branching ratios for the
$\pi^0$ and $\eta$ two-photon decay.  An excess $R_\mathrm{data} >
R_\mathrm{hadron}^\mathrm{Kroll-Wada}$ translates into the fraction of
virtual direct photons according to
\begin{equation}
\frac{\gamma_\mathrm{direct}^{*}}
     {\gamma_\mathrm{incl}^{*}}
=
\frac{R_\mathrm{data} - R_\mathrm{hadron}^\mathrm{Kroll-Wada}}
     {R_\mathrm{direct\; \gamma}^\mathrm{Kroll-Wada} 
      - R_\mathrm{hadron}^\mathrm{Kroll-Wada}} \;.
\end{equation}
The direct-photon $p_\mathrm{T}$ spectrum is then obtained by multiplying 
the inclusive photon spectrum measured with the PHENIX electromagnetic
calorimeters by the ratio 
$\gamma_\mathrm{direct}^{*} / \gamma_\mathrm{incl}^{*}$.

\begin{figure}[t]
\includegraphics[height=.37\textheight]{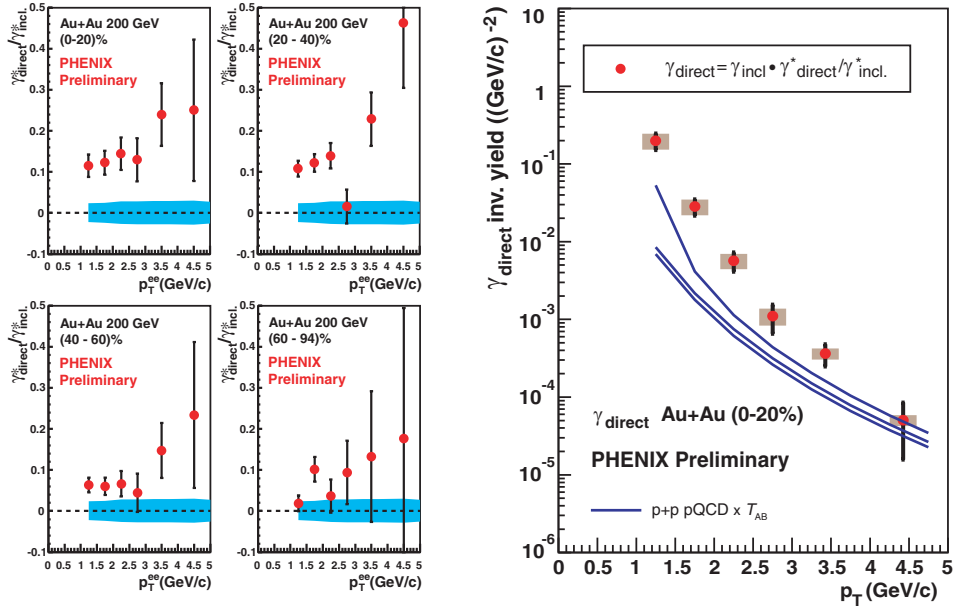}
\caption{Left panel: Ratio of direct to all virtual photons in Au+Au collisions
  at $\sqrt{s_\mathrm{NN}} = 200$\;GeV for four different centrality
  classes ($0-20\,\%$, {\it e.g.}, corresponds to the most central
  class comprising the 20\,\% most central collisions). Right panel:
  Low-$p_\mathrm{T}$ direct-photon spectrum in central Au+Au
  collisions compared to a NLO pQCD calculation.  The three solid
  curves correspond to different scales used in the pQCD calculation
  and indicate the theoretical uncertainties. }
\label{fig:int_conv_result}
\end{figure}

Results from this method are depicted in
Fig.~\ref{fig:int_conv_result}. The left panel shows
$\gamma_\mathrm{direct}^{*} / \gamma_\mathrm{incl}^{*}$ for four
different centrality classes. A significant direct-photon excess in
central Au+Au collisions is observed for $1 \lesssim p_\mathrm{T}
\lesssim 4.5$\;GeV/$c$. The direct-photon excess appears to decrease
towards more peripheral collisions. The systematic uncertainty of
$\gamma_\mathrm{direct}^{*} / \gamma_\mathrm{incl}^{*}$ on the order
of 25\,\% is dominated by the uncertainty of the $\eta/\pi^0$ ratio.
The right panel of Fig.~\ref{fig:int_conv_result} shows the
low-$p_\mathrm{T}$ direct-photon spectrum in central Au+Au collisions
compared to a next-to-leading-order (NLO) p+p pQCD calculation scaled
by the nuclear overlap function $T_{AA}$. The pQCD calculation
provides an estimate of the contribution of hard processes to the
direct-photon spectrum at low $p_\mathrm{T}$. It is planned to replace
the pQCD calculation by direct-photon measurements in p+p and d+Au
collisions in the future.

The excess of the measured direct-photon spectrum in
Fig.~\ref{fig:int_conv_result} over the photons from hard processes
as estimated by the QCD calculation hints at a significant
contribution of thermal photons in central Au+Au collisions at
$\sqrt{s_\mathrm{NN}} = 200$\;GeV. In order to describe thermal photon
production in A+A collisions the entire space-time evolution of the
fireball needs to be modeled.  This is often done with hydrodynamic
calculations which assume local thermal equilibrium. An important free
parameter in such models is the initial temperature of the fireball.
It has been shown in different calculation that the low-$p_\mathrm{T}$
direct-photon spectrum in central Au+Au collisions can be described as
the sum of hard pQCD photons and thermal photons from the QGP and the
hadron gas \cite{Gale:2005zd, Alam:2005za, Chaudhuri:2005dp,
  d'Enterria:2005vz}.  The initial temperatures derived from these
calculation are in the range $370 \lesssim T_\mathrm{i} \lesssim
570$\;MeV. These temperatures are significantly above the critical
temperature for the QGP phase transition of $T_\mathrm{c} \approx
170$\;MeV.

\section{Conclusions}
Direct-photon data from Au+Au collisions at $\sqrt{s_\mathrm{NN}} =
200$\;GeV have been presented for $1 \lesssim p_\mathrm{T} \lesssim
13$\;GeV/$c$. Unlike pions and other hadrons direct photons at high
$p_\mathrm{T}$ are not suppressed, {\it i.e.}, they follow
$T_\mathrm{AA}$ scaling. Thus, the control measurement possible with
high-$p_\mathrm{T}$ direct photons shows that hadron suppression is a
final state effect, consistent with parton energy loss in the QGP.
The low-$p_\mathrm{T}$ direct-photon spectrum in central Au+Au
collisions at $\sqrt{s_\mathrm{NN}} = 200$\;GeV might contain a
significant contribution of thermal photons. Model descriptions of the
data yield initial temperatures above $T_\mathrm{c}$.  This would
provide further evidence for a QGP formation in central Au+Au
collisions at RHIC if the NLO pQCD estimate of the contribution of
direct photons from hard processes at low $p_\mathrm{T}$ is confirmed
by p+p and d+Au measurements.
\bibliography{sample}



\end{document}